\documentclass[aps,pra,twocolumn,superscriptaddress,reprint]{revtex4-2}
\usepackage{float}
\usepackage{graphicx}
\usepackage{amsmath}
\usepackage{bm}
\usepackage{color}
\usepackage{amssymb}
\usepackage{mathrsfs}
\usepackage{dcolumn}
\usepackage{multirow}
\usepackage[colorlinks=true,linkcolor=blue,urlcolor=blue,citecolor=blue]{hyperref}

\begin{document}
	
\title{Intertwined space-time symmetry, orbital magnetism and dynamical Berry connection in a circularly shaken optical lattice}
	
\author{Hua Chen}
\email{Electronic address: hwachanphy@zjnu.edu.cn} 
\affiliation{Department of Physics, Zhejiang Normal University, Jinhua 321004, China}
	
\author{W. Vincent Liu}
\email{Electronic address: wvliu@pitt.edu} 
\affiliation{Department of Physics and Astronomy, University of Pittsburgh, Pittsburgh, Pennsylvania 15260, USA}
\affiliation{Wilczek Quantum Center, School of Physics and Astronomy and T. D. Lee Institute, Shanghai Jiao Tong University, Shanghai 200240, China}		
\affiliation{Shenzhen Institute for Quantum Science and Engineering and Department of Physics, Southern University of Science and Technology, Shenzhen 518055, China}
	
\begin{abstract}
We study the circular shaking of a two dimensional optical lattice, which is essentially a (2+1) dimensional space-time lattice exhibiting periodicities in both spatial and temporal dimensions. The near-resonant optical shaking considered here dynamically couples the low-lying $s$ band and the first excited $p$ bands by transferring a photon of shaking frequency. The intertwined space-time symmetries are further uncovered to elucidate the degeneracy in the spectrum solved with the generalized Bloch-Floquet theorem. Setting the chirality of circular shaking explicitly breaks time reversal symmetry and lifts the degeneracy of $p_\pm = p_x \pm ip_y$ orbitals, leading to the local circulation of orbital magnetism, {\it i.e.} the imbalanced occupation in $p_\pm$ orbitals. Moreover, the dynamics of Berry connection is revealed by the time evolution of the Berry curvature and the polarization, which have physical observable effects in experiments. Interestingly, the dynamics is found characterized by a universal phase shift, governed by the time screw rotational symmetry involving a fractional translation of time. These findings suggest that the present lattice-shaking scheme provides a versatile platform for the investigation of the orbital physics and the symmetry-protected dynamics.
\end{abstract}
	
\date{\today}
	
\maketitle

\section{introduction}
\label{sec:intro}

Floquet driving provides a promising recipe to realize a periodic structure in the extra dimension of time, generalizing the conventional concept of lattice that is spatially periodic. The Floquet shaken optical lattice~\cite{Eckardt7}, which is essentially a space-time lattice with periodic structures in both spatial and temporal dimensions, has been the recent focus of theoretical~\cite{Dalibard11,Goldman14,Goldman16} and experimental~\cite{Jotzu14,Flaschner16,Wintersperger20} research for the purpose of synthetic gauge fields~\cite{Cooper19,Takashi19,Harper20}. From the symmetry perspective, the Berry curvature of a static lattice, which provides indispensable information to characterize the topology of Bloch bands, vanishes strictly across the Brillouin zone if the system preserves both space inversion and time reversal symmetries. However, the Floquet driving generally entwines the space and time dimensions and can break these symmetries, thereby losing the symmetry constraint on the Berry curvature and enriching the topology. For instance, a recent experiment at Munich probed the Berry curvature by measuring the Hall deflection in a Floquet driven honeycomb lattice and identified various types of topological phases~\cite{Wintersperger20}. 
In particular, the anomalous Floquet topological phase they realized has no static counterpart and is characterized by the topology of Bloch-Floquet bands, reflecting the intertwined space-time nature~\cite{Kitagawa10,Levin13,Nathan15}. 
Besides the success of topology, the orbital degree of freedom is also a recent research focus in the Floquet driving. Experimentally, the lattice-shaking scheme with the near-resonant shaking dynamically activates the $p$ orbital, which is originally separated from the low-lying $s$ orbital in the static  lattice~\cite{Gemelke05,Chin13,Chin15,Chicago16,Chin18,Chicago18}. The recent experimental advance leads to a rich variety of interesting phenomena. For instance, a double-well dispersion from the reconstructed band structure through the shaking of one dimensional optical lattice promotes the quantum simulation of ferromagnetism~\cite{Chin13} , roton-maxon excitations in Bose-Einstein condensates~\cite{Chin15}, and the quantum critical dynamics~\cite{Chicago16,Chin18}. Later, the shaking scheme is extended to a two dimensional optical lattice, generalizing the previous experiments in one dimension~\cite{Chicago18}. Theoretically, a quantum phase transition in one dimensional shaken lattices is predicted to exhibit a $\mathbb{Z}_2$ superfluid phase owing to the double-well structure~\cite{Zhai14}. While the generalization to two dimensional shaken lattices is nontrivial, the $p$-orbital angular momentum is quenched in one dimension but is revival in two dimensions. The early studies on static interacting systems have shown that the orbital angular momentum (OAM) arises from the spontaneous time reversal symmetry breaking and renders a nonvanishing Berry curvature, which is the key attribute of emerging exotic quantum states~\cite{Wirth10,SoltanPanahi11,Vincent11,Kock15,Kock16,Li16}. For two dimensional shaken lattices under the near-resonant condition, the theoretical understanding starting from symmetry aspects however remains open due to the activation of $p$ orbitals.

Here we study the (2+1) dimensional space-time lattice that is synthesized by circularly shaking a two dimensional optical lattice with the quasifrequency spectrum being solved with the generalized Bloch-Floquet theorem. Beyond the primitive space and time translational symmetries, the intertwined space-time symmetry, which can not be decomposed as a direct product of spatial and temporal symmetries, is uncovered to further reveal the symmetry-protected degeneracy in the quasifrequency spectrum. We also show that the time reversal symmetry breaking, caused by the chirality of circular shaking, lifts the degeneracy between the dynamical activated $p_\pm = p_x \pm ip_y$ orbitals, resulting in the local circulation of OAM. Moreover, both Berry curvature and polarization, which can be directly derived from the Berry connection, are inherently dynamical due to the circularly shaking. 
The dynamics is further studied by numerically evaluating the time evolution of the Berry curvature and the polarization. Finally, the frequency-domain analysis reveals that the dynamics is characterized by a universal phase shift, originating from the intertwined space-time symmetry.

The rest of paper is organized as follows. In Sec.~\ref{sec:BFS}, we solve the Bloch-Floquet spectrum of the (2+1) dimensional space-time lattice. The intertwined space-time symmetry is discussed in Sec.~\ref{sec:sts} by revealing the symmetry-protected degeneracy in the Bloch-Floquet spectrum. We also discuss the orbital magnetism and the dynamical Berry connection in Secs.~\ref{sec:oam} and \ref{sec:dgf}, respectively. Finally, we summarize our results and discuss the possible generalization in Sec.~\ref{sec:sum}.

\begin{figure}
	\centering
	\includegraphics[width=0.5\textwidth]{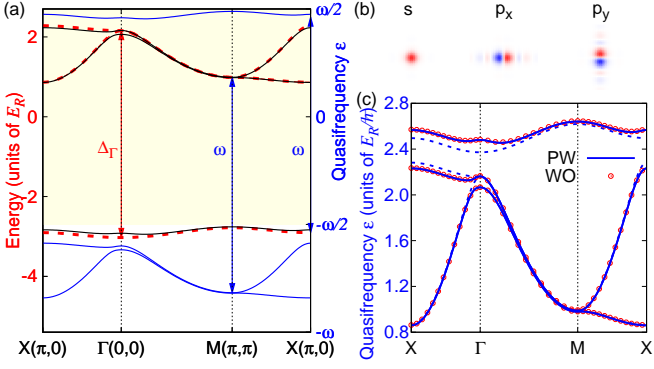}
	\caption{
		(a) Band structure along high symmetry lines in the first Brillouin zone. 
		Red dash lines denote to the lowest three Bloch bands of the static potential $\mathcal{V}_{0}\left(\bm{r}\right)$ in Eq.~(\ref{eq:OLT}). Solid lines are the Bloch-Floquet bands of the shaking optical potential in Eq.~(\ref{eq:SOL}). Black and blue lines trace the original and shifted Bloch-Floquet bands due to the shaking, respectively. The first Floquet Brillouin zone $-\omega/2 < \hbar\epsilon \leqslant \omega/2$ is indicated by yellow shaded region.  
		(b) Spatial distribution of Wannier-orbital wave function for the lowest three Bloch bands. The sign $+(-)$ of wave functions is indicated by red (blue) color. 
		(c) Band structures from plane-wave expansion (blue solid lines) and Wannier-orbital interpolation (open red circles).
		Blue dash lines are shifted Bloch bands from plane-wave expansion for comparison.
		The parameters are $\{V,\omega,\theta_s\}=\{4E_\text{R},5.4E_\text{R},0.1\}$, where the recoil energy is defined as $E_\text{R}=\hbar^2k_\text{L}^2/2m$. The higher orbital bands are not shown since their coupling to $s$ orbital band is negligible for weak shaking.
	}
	\label{fig:band}
\end{figure}

\begin{figure*}
	\centering
	\includegraphics[width=0.98\textwidth]{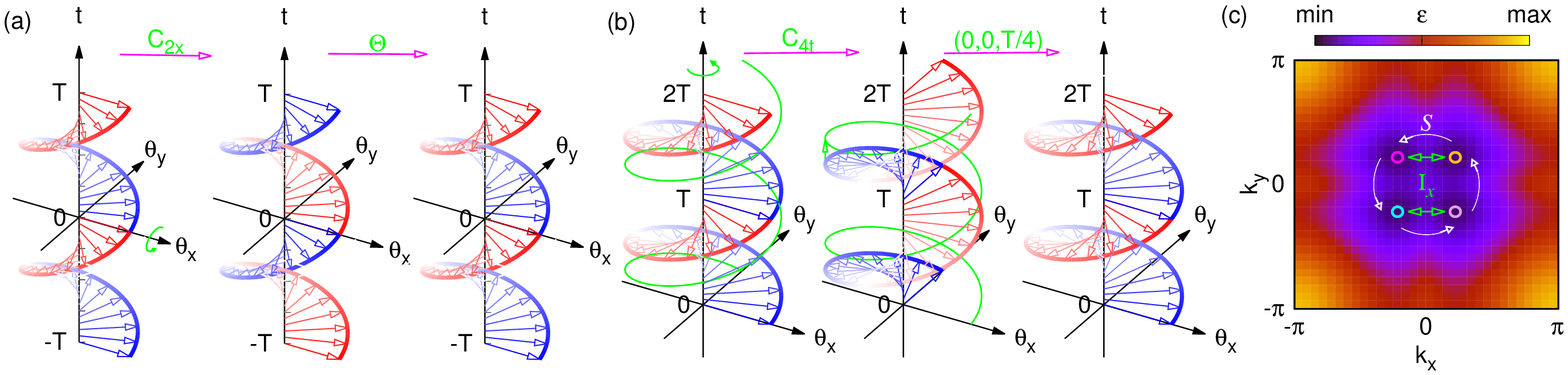}
	\caption{
		Trajectory of shaking phase $\bm{\theta}\left(t\right)=\left(\theta_x\left(t\right),\theta_y\left(t\right)\right)$ in Eq.~(\ref{eq:SOL}) to illustrate (a) intertwined space-time operation $\mathcal{I}_x=\Theta \{C_{2x}|\bm{0}\}$ composed of a space-time rotation of $\pi$ about $x$ axis followed by a time reversal $\Theta$, and (b) time screw rotation $\mathcal{S}=\{C_{4t}|\left(0,0,T/4\right)\}$ composed of a spatial rotation of $\pi/2$ about the temporal axis $t$ followed by a time translation $T/4$.
		(c) Landscape of the quasifrequency $\epsilon$ for the $s$ Bloch-Floquet band. Four degenerate band minima (valleys), indicated by the open circles with different colors, illustrate the symmetry-protected degeneracy between the Bloch-Floquet states at different quasimomenta. The parameters are chosen as $\{V,\omega,\theta_s\}=\{4E_\text{R},5.4E_\text{R},0.1\}$.  
	}
	\label{fig:sym}
\end{figure*}

\section{Bloch-Floquet spectrum}
\label{sec:BFS}

We start with the following (2+1) dimensional optical potential
\begin{eqnarray}
\mathcal{V}\left(\bm{r},t\right) = -V\{
 \cos\left[qx-\theta_x\left(t\right)\right]
&+&\cos\left[qy-\theta_y\left(t\right)\right]
\}, \nonumber\\
\{\theta_x\left(t\right),\theta_y\left(t\right)\}
= \theta_s\{
\cos\left[\omega t\right],
&\sin&\left[\omega t\right]
\},\text{ }q=2k_\text{L}
\label{eq:SOL}
\end{eqnarray}
with $k_\text{L}$ being the wave vector of laser beams. It describes that a two dimensional square optical lattice with lattice depth $V$ and lattice spacing $d=\pi/k_\text{L}$ is circularly shaken with amplitude $\theta_s$ and frequency $\omega$~\cite{Chicago18}. Without loss of generality, we focus on the shaking with right circular polarization $\bm{\theta}=\left(\theta_x,\theta_y\right)$ in the (2+1) dimensional space-time coordinate. Different chiralities of shaking polarizations are simply connected by a space mirror operation. The optical potential $\mathcal{V}\left(\bm{r},t\right)$ in Eq.~(\ref{eq:SOL}) is invariant under discrete translations with integer multiples of space and time primitive vectors $\{d\hat{x},d\hat{y},T\hat{t}\}$ where $T=2\pi/\omega$ is temporal periodicity. The time-dependent Schr\"odinger equation $i\hbar\partial_t\psi\left(\bm{r},t\right)=\mathcal{\hat{H}}\left(\bm{r},t\right)\psi\left(\bm{r},t\right)$ 
with $\mathcal{\hat{H}}\left(\bm{r},t\right)=\hat{\bm{p}}^2/2m+\mathcal{V}\left(\bm{r},t\right)$ can be solved by using the generalized Bloch-Floquet theorem with its $\mu$-th eigenstate $\psi_{\mu\bm{k}}\left(\bm{r},t\right)=\exp\left[i\left({\bm{k}}\cdot\bm{r}-\epsilon_{\mu\bm{k}} t\right)\right]u_{\mu\bm{k}}\left(\bm{r},t\right)$ labeled by quasimomentum $\bm{k}$ and quasifrequency $\epsilon_{\mu\bm{k}}$. The Bloch-Floquet orbital wavefunction inherits the periodicities of space-time optical potential $\mathcal{V}\left(\bm{r},t\right)$ and takes the form
\begin{equation}
	u_{\mu\bm{k}}\left(\bm{r},t\right)
	=\sum_{{\bf{G}}\Omega}\exp\left[i\left({\bf{G}}\cdot\bm{r}-\Omega t\right)\right]u_{\mu\bm{k}}^{{\bf{G}}\Omega} 
\end{equation}
with ${\bf G}$ and $\Omega$ being the reciprocal space and time vectors, respectively.
Following the Floquet theory, the time-dependent Schr\"odinger equation is then converted into an eigenvalue problem 
\begin{equation}
	\sum_{{\bf{G}}^\prime\Omega^\prime}\hat{K}_{{\bf G}\Omega,{\bf G}^\prime\Omega^\prime}
	u_{\mu\bm{k}}^{{\bf G}^\prime\Omega^\prime}=\hbar\epsilon_{\mu\bm{k}} u_{\mu\bm{k}}^{{\bf G}\Omega} 
\end{equation}
with 
\begin{equation}
\hat{K}_{{\bf G}\Omega,{\bf G}^\prime\Omega^\prime}
=\left[\frac{\left(\hbar\bm{k}+\hbar{\bf G}\right)^2}{2m}-\hbar\Omega\right]
\delta_{{\bf G}\Omega,{\bf G}^\prime\Omega^\prime}
+\mathcal{V}_{{\bf G}-{\bf G}^\prime\Omega-\Omega^\prime} 
\end{equation}
being the Floquet operator $\hat{K}\left(\bm{r},t\right)=\hat{\mathcal{H}}\left(\bm{r},t\right)-i\hbar\partial_t$ 
in the Fourier-transformed bases~\cite{Shirley65,Sambe73}. 
Here $\mathcal{V}_{{\bf G}\Omega}$ are the Fourier component of the space-time potential $\mathcal{V}\left(\bm{r},t\right)$. 
Before proceeding, it is instructive to discuss the optical potential 
in temporal series $\mathcal{V}\left(\bm{r},t\right)=\sum_n \mathcal{V}_n\left(\bm{r}\right)\exp\left[-in\omega t\right]$
with
\begin{eqnarray}
\mathcal{V}_{n}\left(\bm{r}\right) = -V_n\times 
\left \{
\begin{tabular}{cc}
$-i^{n+1}\sin\left(qx\right)+i\sin\left(qy\right)$, & $n$ odd  \\
$i^n\cos\left(qx\right)+\cos\left(qy\right)$, & $n$ even 
\end{tabular}
\right. 
\label{eq:OLT}
\end{eqnarray}
where $V_n = VJ_n\left(\theta_s\right)$ is the effective potential and $J_n(x)$ is the Bessel function of first kind. The zeroth-order optical potential $\mathcal{V}_{n=0}\left(\bm{r}\right)$ gives rise to a static square optical potential with effective potential $V_0=VJ_0\left(\theta_s\right)$. The corresponding Bloch band structure is shown in Fig.~\ref{fig:band}(a).
For the lowest three bands, Wannier orbitals are further constructed to reveal the band characters
\begin{equation}
w_{\mu {\bf R}}\left(\bm{r}\right) = \sum_{\bm{k}}\exp\left[i\bm{k}\cdot{\bf R}\right]\bar{\psi}_{\mu\bm{k}}\left(\bm{r}\right),
\label{eq:Wannier}
\end{equation}
where ${\bf R}$ specifies the center of Wannier orbitals and $\bar{\psi}_{\mu\bm{k}}\left(\bm{r}\right)$ is the $\mu$-th Bloch state of the static potential $\mathcal{V}_{n=0}\left(\bm{r}\right)$ in Eq.~(\ref{eq:OLT}) in the Wannier gauge. Technical details are presented in Appendix~\ref{app:WO}. The lowest and first excited two bands are characterized by $s$ and $\left(p_x,p_y\right)$ orbitals, respectively. Their spatial distributions of wave functions are depicted in Fig.~\ref{fig:band}(b). 
Upon shaking, the Bloch bands are repeatedly shifted by shaking frequency $\omega$ and are hybridized by the dynamical optical potential $\mathcal{V}_{n\ne0}\left(\bm{r}\right)$. As sketched in Fig.~\ref{fig:band}(a), we focus on the blue detuned case $\omega\gtrsim \Delta_\Gamma$ such that the shaking frequency $\omega$ is slightly higher than the direct band gap $\Delta_\Gamma$ between $s$ and $\left(p_x, p_y\right)$ bands at $\Gamma$ point. In the weak shaking limit, the effective potential of high order $V_n$ is severely suppressed by the asymptotic behavior of Bessel function $J_{\left|n\right|}(x)\approx\left(x/2\right)^{\left|n\right|}/\Gamma\left(\left|n\right|+1\right)$, 
where $\Gamma\left(x\right)$ is the gamma function. In practice, the optical potential $\mathcal{V}_n\left(\bm{r}\right)$ in our numerical simulations are truncated up to $\left|n\right|=1$. As shown in Fig.~\ref{fig:band}(c), the $s$ and $\left(p_x, p_y\right)$ bands are hybridized by the dynamical optical potential $\mathcal{V}_{n=\pm1}\left(\bm{r}\right)$, accompanied by absorbing or emitting a photon of shaking frequency. The Bloch-Floquet band structures calculated by plane-wave expansion and Wannier-orbital interpolation show excellent agreement with each other.

\section{intertwined space-time symmetry}
\label{sec:sts}

Having established the Bloch-Floquet band structure, we next show that the present (2+1) dimensional system possesses a $D_4$ space-time symmetry. Besides the primitive spatial and temporal translational symmetry discussed above, the intertwined space-time symmetries are essential to understand the degeneracy of the quasifrequency spectrum at different quasimomenta connected by the symmetry operation. To facilitate discussion, we adopt a (2+1)-dimensional space-time vector $\left(\bm{r},t\right)$ to define a space-time operation
\begin{equation} 
\{g|\bm{\tau}\}\left(\bm{r},t\right)=g\left(\bm{r},t\right)+\bm{{\tau}}, 
\end{equation}
where $g$ and $\bm{\tau}$ denote point group operations and translational vectors, respectively.
The $D_4$ space-time group has two group generators in total. 
As depicted in Fig.~\ref{fig:sym}(a), the first generator is 
\begin{eqnarray}
	\mathcal{I}_x=\Theta \{C_{2x}|\bm{0}\}, 
\end{eqnarray}
which is a space-time rotation of $\pi$ about $x$ axis that sends the space-time vector $\left(\bm{r},t\right)$ to $\left(x,-y,-t\right)$, followed by a time reversal $\Theta$. For spinless bosons, the time reversal operator is mathematically described by complex conjugate $\Theta=\mathcal{K}$. It is worth remarking that the symmetry discussed here refers to the space-time transformation $\hat{U}\left(\bm{r},t\right)$ under which the Floquet operator $\hat{K}=\hat{\mathcal{H}}-i\hbar\partial_t$ is left invariant. As depicted in Fig.~\ref{fig:sym}(b), the second generator of $D_4$ space-time group is the time screw rotation 
\begin{eqnarray}
\mathcal{S}=\{C_{4t}|\left(0,0,T/4\right)\}.	
\end{eqnarray}
It describes a spatial rotation of $\pi/2$ about the temporal axis $t$, which sends the space-time vector $\left(\bm{r},t\right)$ to $\left(-y,x,t\right)$, followed by a fractional primitive translation of time $T/4$. These two symmetries have crucial implications on the Bloch-Floquet band structure. To illustrate, we focus on the $s$ Bloch-Floquet band that is adiabatically connected to the $s$ band in the non-shaking limit. The symmetry prediction is however not limited to this band. As shown in Fig.~\ref{fig:sym}(c), the intertwined space-time symmetry $\mathcal{I}_x$ ensures an identical spectrum at quasimomenta $\left(k_x,k_y\right)$ and $\left(-k_x,k_y\right)$ by constructing an explicit connection between their Bloch-Floquet orbital wavefunctions. Similarly, the time screw rotational symmetry $\mathcal{S}$ connects the states at $\left(k_x,k_y\right)$ and $\left(-k_y,k_x\right)$. Detailed proofs are provided in Appendix~\ref{app:sym}. Particular attention should be paid to four degenerated band minima (dubbed valleys), which are fully connected by the space-time symmetries. These valleys arise from the dynamical hybridization of $s$ and $\left(p_x, p_y\right)$ orbitals due to lattice shaking, and are generally incommensurate to the optical lattice. The single-particle ground state, constructed by the linear superposition of the wave functions at these valleys, has an infinite degeneracy. This single-particle degeneracy can be lifted by many-body interactions through the spontaneous symmetry breaking. In the weakly interacting limit, the interacting energy is further estimated within the Gross-Pitaevskii approximation~\cite{Dalfovo99,pethick08}. The ground state is found to be valley polarized. This is different from the commensurate case that the intervalley exchange interaction promoted by the umklapp scattering process supports the valley coherent Bose-Einstein condensation~\cite{Chen20,nozieres95}. 

\begin{figure}
	\centering
	\includegraphics[width=0.5\textwidth]{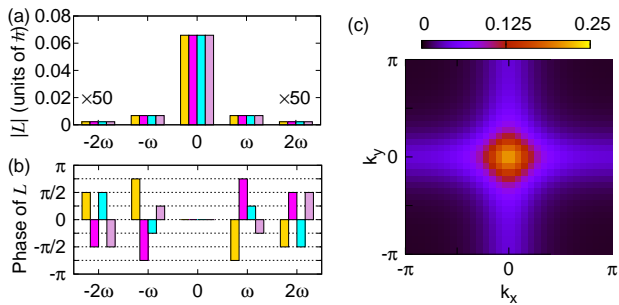}
	\caption{ (a) Amplitude and (b) phase of the complex-valued local circulation of orbital angular momentum 
		$L\left(\bm{k},\Omega\right)$ in Eq.~(\ref{eq:Lkomega}) at four valleys with the representation of bar colors indicated in Fig.~\ref{fig:sym}(c). (c) Static orbital angular momentum $L\left(\bm{k},\Omega=0\right)$ in the first Brillouin zone for $s$ Bloch-Floquet band.
		The parameters in numerical calculations are $\{V,\omega,\theta_s\}=\{4E_\text{R},5.4E_\text{R},0.1\}$.
	}
	\label{fig:oam}
\end{figure}

\section{orbital magnetism}
\label{sec:oam}

The chirality of shaking polarization in Eq.~(\ref{eq:SOL}) explicitly breaks the time-reversal symmetry, which leads to the generation of orbital magnetism, {\it i.e.} the imbalanced occupation in $p_{\pm}=p_x \pm i p_y$ orbitals. 
To gain an intuitive understanding, we take a close inspection on the optical potential Eq.~(\ref{eq:OLT}) in temporal series.
In polar coordinates $\left(r,\theta\right)$, the leading order expanded in terms of partial waves takes the form
\begin{equation}
	\mathcal{V}_{n=\pm 1}\left(\bm{r}\right)= - 2V_{\pm 1} \sum_{\ell \text{ odd}}
	\Xi_\pm^\ell
	J_\ell\left(qr\right)\exp\left[i\ell\theta\right]
\end{equation}
with the symmetry indicator $\Xi_\pm^\ell=\pm\left(i^{\ell-1}\pm1\right)/2$.
The transition probabilities between the local $s$ and $p_\pm$ Wannier orbitals are exclusively determined by the potentials with partial waves $\ell=\pm1$ according to the selection rule of OAM~\cite{Chen18}.
Furthermore, the time reversal symmetry breaking induced by the circular shaking is indicated by $\Xi_\pm^\ell$. Specifically, the transition from the $s$ orbital to the $p_+$ ($p_-$) orbital by absorbing a photon is permitted $\Xi_+^{+1}=1$ (forbidden $\Xi_+^{-1}=0$). In contrast, the transition from the $p_+$ ($p_-$) orbital to the $s$ orbital by emitting a photon is forbidden $\Xi_-^{+1}=0$ (permitted $\Xi_-^{-1}=1$). Consequently, the degeneracy lifting between the $p_{\pm}$ Wannier orbitals leads the local circulation of OAM around the lattice sites in the bulk, which is extensively studied in electronic solids in the context of orbital magnetism~\cite{Xiao05,Thonhauser05,Ceresoli06,Shi07,Xiao10,Thonhauser11}. Here, we extend the microscopic derivation on the local circulation of OAM by explicitly calculating the contribution of a single Bloch-Floquet state, which is crucial for the valley polarized Bose-Einstein condensation. Importantly, the OAM studied here is inherently dynamical due to the lattice shaking, which sets our study apart from the previous studies on static lattices. The contribution of $s$-band Bloch-Floquet state $\psi_{s\bm{k}}\left(\bm{r},t\right)$ in frequency domain is mathematically described by 
\begin{equation}
	L\left(\bm{k},\Omega\right) = \frac{1}{\mathcal{N}_\text{uc}}\sum_{\bf R}
	\frac{1}{T}\int_0^T dt \left\langle\psi_{s\bm{k}}|
	\hat{L}^z_{\bf R}
	|\psi_{s\bm{k}}\right\rangle
	\exp\left[i\Omega t\right], 
	\label{eq:Lkomega}
\end{equation}
where $\hat{\bm{L}}_{\bf R}=\left(\bm{r}-{\bf R}\right)\times\hat{\bm{p}}$ is the OAM operator around the static lattice sites ${\bf R}$ and $\mathcal{N}_\text{uc}$ is the number of unit cells. A lengthy but straightforward algebra leads to the explicit expression
\begin{eqnarray}
&	L&\left(\bm{k},\Omega^\prime-\Omega\right)=-i\hbar\sum_{{\bf{G}\ne\bf{G}^\prime}}\sum_{\Omega\Omega^\prime}
	u_{s\bm{k}}^{{\bf G}\Omega *}u_{s\bm{k}}^{{\bf G}^\prime\Omega^\prime}\nonumber\\
&\times&	\{\exp\left[-i\left(G_x^\prime-G_x\right)\frac{d}{2}\right]
	\frac{k_y+G_y^\prime}{G_x^\prime-G_x}\delta_{G_y,G_y^\prime}-x\leftrightarrow y\}.\nonumber
\end{eqnarray}
Figures~\ref{fig:oam}(a) and~\ref{fig:oam}(b) show the numerical evaluation of $L\left(\bm{k},\Omega\right)$ at four valleys. The static OAM dominates while the high frequency component at $\Omega=\pm2\omega$ is vanishingly small. The phase of dynamical OAM at neighboring valleys differs by $\pi/2$ and $\pi$ for $\Omega=\pm\omega$ and $\pm2\omega$, respectively. This phase shift between neighboring valleys is dictated by the fractional translation of time $T/4$ in the time screw rotation symmetry $\mathcal{S}$ and thus serves as a signature of the intertwined space-time symmetry. The static OAM in the first Brillouin zone shown Fig.~\ref{fig:oam}(c) reaches its maximum at $\Gamma$ point where the coupling between $s$ and $p_+$ orbitals is most significant.
Remarkably, the emergent local circulation of OAM under the near-resonant condition distinguishes our study from the early study, in which an additional phase of inter-site tunneling is imprinted by Floquet driving~\cite{Jotzu14}.

\begin{figure}
	\centering
	\includegraphics[width=0.5\textwidth]{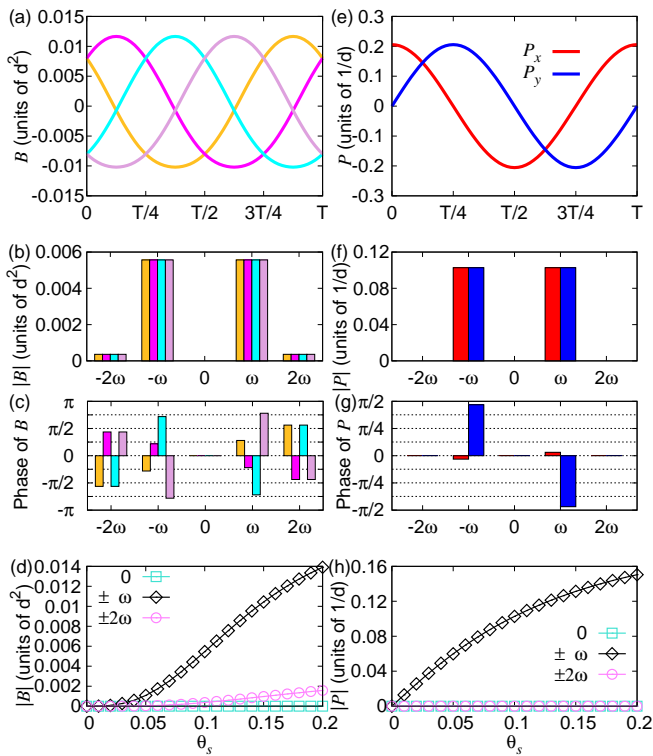}
	\caption{ (a) Time evolution of Berry curvature $\mathcal{B}\left(\bm{k},t\right)$ in Eq.~(\ref{eq:berry}), (b) amplitude and (c) phase of the complex-valued Berry curvature $\mathcal{B}\left(\bm{k},\Omega\right)$ in Eq.~(\ref{eq:berryfft}) at four valleys with the representation of colors indicated in Fig.~\ref{fig:sym}(c). (e) Time evolution of dynamical polarization $\bm{\mathcal{P}}\left(t\right)$ in Eq.~(\ref{eq:polarization}), (f) amplitude and (g) phase of the complex-valued polarization $\bm{\mathcal{P}}\left(\Omega\right)$ in Eq.~(\ref{eq:polarizationfft}) for the $s$ Bloch-Floquet band. The Berry curvature $|\mathcal{B}|$ among different valleys and the polarization $|\bm{\mathcal{P}}|$ along different directions in (b) and (f) share an identical amplitude at the same frequency, respectively. The frequency-resolved identical amplitudes of (d) $|\mathcal{B}|$ and (h) $|\bm{\mathcal{P}}|$ as a function of shaking amplitude $\theta_s$. The parameters are taken as $\{V,\omega,\theta_s\}=\{4E_\text{R},5.4E_\text{R},0.1\}$. The shaking amplitude $\theta_s$ in (d) and (h) is varied to study its dependence.
	}
	\label{fig:berry}
\end{figure}

\section{dynamical Berry connection}
\label{sec:dgf}

The Berry phase of the $s$ Bloch-Floquet band, which is inherently dynamical due to the circular shaking, can be expressed as an integral of the Berry connection
\begin{equation}
	\mathcal{A}_{\bm{k}}\left(t\right)=i\left\langle u_{s\bm{k}}\left(\bm{r},t\right)|\nabla_{\bm{k}}|u_{s\bm{k}}\left(\bm{r},t\right)\right\rangle
\end{equation}
over a closed path. Mathematically, the Berry curvature describing the local Berry phase can be defined in terms of the Berry connection via the Stokes theorem
\begin{equation}
	\mathcal{B}\left(\bm{k},t\right)=\nabla_{\bm{k}}\times{\mathcal A}_{\bm{k}}\left(t\right),
	\label{eq:berry}
\end{equation}
which is a gauge invariant quantity with physical significance~\cite{Bohm03,Wilczek89}.
The numerical evaluation of Berry curvature is based on an efficient method in discretized Brillouin zone~\cite{Takahiro05}. 
As depicted in Fig.~\ref{fig:berry}(a), the time evolution of Berry curvature $\mathcal{B}\left(\bm{k},t\right)$ at four valleys shows an oscillating behavior. For the valley polarized Bose-Einstein condensation, the dynamical Berry curvature can be in principal detected by measuring the transverse shift of anomalous velocity $\bm{\mathcal{F}}\times\mathcal{B}\left(\bm{k},t\right)$ in the presence of an external force $\bm{\mathcal{F}}$, which can be engineered by a slowly varying optical potential~\cite{Goldman13,Liu13,Aidelsburger15}. This transverse shift resembles that of charged particles under ac magnetic fields. The dynamics is quantitatively characterized by introducing the Fourier-transformed Berry curvature
\begin{equation}
	\mathcal{B}\left(\bm{k},\Omega\right)=	\frac{1}{T}\int_{0}^{T}dt\mathcal{B}\left(\bm{k},t\right)\exp\left[i\Omega t\right]
	\label{eq:berryfft}
\end{equation}
in the frequency domain. As shown in Fig.~\ref{fig:berry}(b), the dynamical Berry curvature is dominated at the frequency $\Omega=\pm\omega$ with a vanished static component, which is in stark contrast with the dynamics of the local circulation of OAM. The phase shift of Berry curvature between neighboring valleys in Fig.~\ref{fig:berry}(c) follows the prediction of intertwined space-time symmetry discussed above. Figure~\ref{fig:berry}(d) shows that the Berry curvature in frequency domain grows monotonically as the shaking amplitude increases. Since the Berry curvature is purely dynamical and generally weak, the experimental detection through measuring the transverse shift of the micromotion within a driving period may be challenging. 
Alternatively, the dynamics of Berry connection can be inferred from the dynamical polarization
\begin{equation}
	\bm{\mathcal{P}}\left(t\right) = \frac{1}{2\pi}\int_\text{BZ} d\bm{k}\mathcal{A}_{\bm{k}}\left(t\right), 
	\label{eq:polarization}
\end{equation} 
which can be derived from the dynamical Berry connection through the Wannier-Bloch duality between real and momentum space~\cite{Resta2007,Vanderbilt18}.
The dynamical polarization $\mathcal{P}_{\mu}\left(t\right)$ corresponds to the averaged Berry phase accumulated along the closed path in the direction $k_\mu$, which is also proportional to the shift of the instantaneous Wannier-orbital center away from the neighboring lattice site in direction $\mu$~\cite{Liu20}. 
Figure~\ref{fig:berry}(e) shows the numerical results on the time evolution of the dynamical polarization, which is locked in phase with the circular driving. Similarly, the frequency-domain analysis applies to the dynamical polarization
\begin{equation}
	\bm{\mathcal{P}}\left(\Omega\right) = \frac{1}{T}\int_0^Tdt\bm{\mathcal{P}}\left(t\right)\exp\left[i\Omega t\right]. 
	\label{eq:polarizationfft}
\end{equation}
Figures~\ref{fig:berry}(f) and \ref{fig:berry}(g) show that the dynamical polarization oscillates at frequency $\Omega=\pm\omega$ with a universal $\pi/2$ phase shift between $\mathcal{P}_x\left(\pm\omega\right)$ and $\mathcal{P}_y\left(\pm\omega\right)$. As shown in Fig.~\ref{fig:berry}(h), the shaking amplitude generally enhances the frequency resolved polarization as expected.
Experimentally, the dynamical polarization can be determined by measuring the center of mass position of ultracold atoms through  {\it in situ} imaging. For instance, such a measurement was done in a dynamically controlled optical superlattice~\cite{Nakajima16,Lohse16}.

\section{summary and outlook}
\label{sec:sum}
To summarize, we have studied the intertwined space-time symmetries as well as their implication on dynamics for the circularly shaken optical lattice. In particular, the time screw rotation symmetry studied here can be easily generalized to the $n$-fold spatial rotational axis in the presence of a circularly polarized ac field. The symmetry-protected phase shift is generally applicable as a universal feature of the dynamical response for these systems. Moreover, the orbital angular momentum is shown as a direct consequence of time reversal symmetry breaking induced by the circular shaking. Finally, the dynamical nature of Berry connection is revealed by studying the time evolution of the Berry curvature and the polarization. Our study mainly focusing on the symmetry aspect serves as a starting point for future studies, {\it e.g.} out-of-equilibrium with dissipation.

\section*{Acknowledgments}
We thank Cheng Chin, Logan W. Clark, and Kai-Xuan Yao for helpful discussions. 
This work is supported by National Natural Science Foundation of China under Grants No. 11704338 and No. 11835011 (H. C.), and by AFOSR Grant No.  FA9550-16-1-0006, MURI-ARO Grant No. W911NF-17-1-0323 through UC Santa Barbara, and Shanghai Municipal Science and Technology Major Project (Grant No. 2019SHZDZX01) (W.V. L.).

\appendix

\section{Wannier orbital construction}
\label{app:WO}

The approach of Wannier orbitals for the solid-state materials has relied on the well established Marzari-Vanderbilt scheme~\cite{Vanderbilt97,Vanderbilt01,Vanderbilt12}, and is numerically implemented into a numerical package~\cite{Mostofi08}. The Wannier states are determined by numerically minimizing the spatial spread functional  to find the optimal unitary transformation into the Wannier basis. Here, we adopt an alternative numerical method based on Refs.~\cite{Kivelson82,Resta98} to calculate Wannier states for the static optical $\mathcal{V}_{n=0}\left(\bm{r}\right)$ in Eq.~(\ref{eq:OLT}), reducing the problem to the diagonalization of a matrix. 
This intrinsically avoids the possible problem of trapping  in local minima during the numerical minimization and is numerically highly efficient. 

Following Refs.~\cite{Kivelson82,Resta98}, the Wannier orbitals are the eigenstates of the Resta's position operator $\hat{R}=\exp\left[-i\Delta\bm{k}\cdot\bm{r}\right]$, which satisfies the periodic boundary condition.
In the Bloch band basis, the matrix elements of the operator $\hat{R}$ can be expressed as the real-space integrals over the entire spatial region
\begin{eqnarray}
	R_{\bm{k}\bm{k}^\prime}^{\mu\mu^\prime}=\frac{1}{L_xL_y}\int d\bm{r}\psi_{\mu\bm{k}}^*\left(\bm{r}\right)\hat{R}\psi_{\mu^\prime\bm{k}^\prime}\left(\bm{r}\right),
\end{eqnarray}
where $\psi_{\mu\bm{k}}\left(\bm{r}\right)$ are the Bloch states of the static potential.
The real-space integration can be calculated analytically. Then, the explicit expression reads 
\begin{eqnarray}
	R_{\bm{k}\bm{k}^\prime}^{\mu\mu^\prime}=
	\sum_{{\bf{G}\bf{G}}^\prime}\delta_{\bm{k}+{\bf G}+\Delta\bm{k},\bm{k}^\prime+{\bf G}^\prime}
	u_{\mu\bm{k}}^{{\bf G}*}u_{\mu^\prime\bm{k}^\prime}^{{\bf G}^\prime},
	\label{eq:Rmat}
\end{eqnarray} 
where $u_{\mu\bm{k}}^{\bf G}$ are the Bloch orbitals of the static potential.
Moreover, the separability of the static optical potential
\begin{equation}
	\mathcal{V}_{n=0}\left(\bm{r}\right)=-V_0\left[\cos\left(qx\right)+\cos\left(qy\right)\right]
\end{equation}
ensures the commutation relation of the position operators $[\hat{X},\hat{Y}]=0$. Therefore, the two-dimensional Wannier orbitals can be obtained simultaneously by successively diagonalizing the corresponding matrices $X$ and $Y$ in Eq.~(\ref{eq:Rmat}). 

\section{Intertwined space-time symmetry}
\label{app:sym}

In the following, we provide the detailed proof of the implication of the intertwined space-time symmetry. With the generalized Bloch-Floquet theorem, the time-dependent Schr\"odinger equation is recast to the following equation 
\begin{equation}
	\hat{K}\left(\bm{r},t\right)\exp\left[i\bm{k}\cdot\bm{r}\right]u_{\mu\bm{k}}\left(\bm{r},t\right)
	=\hbar\epsilon_{\mu\bm{k}}\exp\left[i\bm{k}\cdot\bm{r}\right]u_{\mu\bm{k}}\left(\bm{r},t\right)
	\label{eq:Flo}
\end{equation}
with the Floquet operator
\begin{equation}
	\hat{K}\left(\bm{r},t\right)=\hat{\mathcal{H}}\left(\bm{r},t\right)-i\hbar\partial_t.
\end{equation}
First, we shall focus on the time screw rotation
\begin{equation}
	\mathcal{S}\left(x,y,t\right)=\left(-y,x,t+\frac{T}{4}\right).	
\end{equation}
It is easy to check that the Floquet operator is invariant under the time screw rotation 
$\mathcal{S}^\dagger\mathcal{K}\left(\bm{r},t\right)\mathcal{S}=\mathcal{K}\left(\bm{r},t\right)$, 
which ensures that both $\mathcal{S}\exp\left[i\bm{k}\cdot\bm{r}\right]u_{\mu\bm{k}}\left(\bm{r},t\right)$ and $\exp\left[i\bm{k}\cdot\bm{r}\right]u_{\mu\bm{k}}\left(\bm{r},t\right)$ are the eigenstates of Eq.~(\ref{eq:Flo}) and share an identical quasifrequency. On the other hand, the simple relation
\begin{equation}
	\mathcal{S}\exp\left[i\bm{k}\cdot\bm{r}\right]u_{\mu\bm{k}}\left(\bm{r},t\right)=\exp\left[i\left(-k_y,k_x\right)\cdot\bm{r}\right]\mathcal{S}u_{\mu\bm{k}}\left(\bm{r},t\right)
\end{equation}
indicates that $\mathcal{S}u_{\mu\bm{k}}\left(\bm{r},t\right)$ is the eigenstate of Eq.~(\ref{eq:Flo}) with quasimomentum $\left(-k_y,k_x\right)$. Therefore, the degeneracy between $\left(-k_y,k_x\right)$ and $\left(k_x,k_y\right)$ is established with their eigenstates obeying
\begin{equation}
	u_{\mu\left(-k_y,k_x\right)}^{\left(-G_y,G_x\right)\Omega}\propto u_{\mu\bm{k}}^{{\bf G}\Omega}\exp\left[\frac{1}{4}i\Omega T\right].
	\label{eq:twist}
\end{equation}
It is worth remarking that the phase shift discussed in Secs.~\ref{sec:oam} and \ref{sec:dgf} is dictated by the twist phase $\exp\left[i\Omega T/4\right]$ in Eq.~(\ref{eq:twist}).

Second, we would like to briefly discuss the intertwined space-time symmetry 
\begin{equation}
	\mathcal{I}_x=\Theta\{C_{2x}|\bf{0}\}.
\end{equation}
Similar analysis shows that the states $\mathcal{I}_xu_{\mu\bm{k}}\left(\bm{r},t\right)$ and $u_{\mu\bm{k}}\left(\bm{r},t\right)$ share
an identical quasifrequency with the following relation on their Fourier components
\begin{equation}
	u_{\mu\left(-k_x,k_y\right)}^{\left(-G_y,G_x\right)\Omega}\propto u_{\mu\bm{k}}^{{\bf G}\Omega*}
\end{equation}
between $\left(-k_x,k_y\right)$ and $\left(k_x,k_y\right)$ in momentum space. 

It is worth remarking that these results have the origin in symmetry and are independent on detailed Wannier-orbital projections.


\begin{thebibliography}{56}%
	\makeatletter
	\providecommand \@ifxundefined [1]{%
		\@ifx{#1\undefined}
	}%
	\providecommand \@ifnum [1]{%
		\ifnum #1\expandafter \@firstoftwo
		\else \expandafter \@secondoftwo
		\fi
	}%
	\providecommand \@ifx [1]{%
		\ifx #1\expandafter \@firstoftwo
		\else \expandafter \@secondoftwo
		\fi
	}%
	\providecommand \natexlab [1]{#1}%
	\providecommand \enquote  [1]{``#1''}%
	\providecommand \bibnamefont  [1]{#1}%
	\providecommand \bibfnamefont [1]{#1}%
	\providecommand \citenamefont [1]{#1}%
	\providecommand \href@noop [0]{\@secondoftwo}%
	\providecommand \href [0]{\begingroup \@sanitize@url \@href}%
	\providecommand \@href[1]{\@@startlink{#1}\@@href}%
	\providecommand \@@href[1]{\endgroup#1\@@endlink}%
	\providecommand \@sanitize@url [0]{\catcode `\\12\catcode `\$12\catcode
		`\&12\catcode `\#12\catcode `\^12\catcode `\_12\catcode `\%12\relax}%
	\providecommand \@@startlink[1]{}%
	\providecommand \@@endlink[0]{}%
	\providecommand \url  [0]{\begingroup\@sanitize@url \@url }%
	\providecommand \@url [1]{\endgroup\@href {#1}{\urlprefix }}%
	\providecommand \urlprefix  [0]{URL }%
	\providecommand \Eprint [0]{\href }%
	\providecommand \doibase [0]{https://doi.org/}%
	\providecommand \selectlanguage [0]{\@gobble}%
	\providecommand \bibinfo  [0]{\@secondoftwo}%
	\providecommand \bibfield  [0]{\@secondoftwo}%
	\providecommand \translation [1]{[#1]}%
	\providecommand \BibitemOpen [0]{}%
	\providecommand \bibitemStop [0]{}%
	\providecommand \bibitemNoStop [0]{.\EOS\space}%
	\providecommand \EOS [0]{\spacefactor3000\relax}%
	\providecommand \BibitemShut  [1]{\csname bibitem#1\endcsname}%
	\let\auto@bib@innerbib\@empty
	\bibitem [{\citenamefont {Eckardt}(2017)}]{Eckardt7}%
	\BibitemOpen
	\bibfield  {author} {\bibinfo {author} {\bibfnamefont {A.}~\bibnamefont
			{Eckardt}},\ }\href {https://doi.org/10.1103/RevModPhys.89.011004} {\bibfield
		{journal} {\bibinfo  {journal} {Rev. Mod. Phys.}\ }\textbf {\bibinfo
			{volume} {89}},\ \bibinfo {pages} {011004} (\bibinfo {year}
		{2017})}\BibitemShut {NoStop}%
	\bibitem [{\citenamefont {Dalibard}\ \emph {et~al.}(2011)\citenamefont
		{Dalibard}, \citenamefont {Gerbier}, \citenamefont {Juzeliunas},\ and\
		\citenamefont {\"Ohberg}}]{Dalibard11}%
	\BibitemOpen
	\bibfield  {author} {\bibinfo {author} {\bibfnamefont {J.}~\bibnamefont
			{Dalibard}}, \bibinfo {author} {\bibfnamefont {F.}~\bibnamefont {Gerbier}},
		\bibinfo {author} {\bibfnamefont {G.}~\bibnamefont {Juzeliunas}},\ and\
		\bibinfo {author} {\bibfnamefont {P.}~\bibnamefont {\"Ohberg}},\ }\href
	{https://doi.org/10.1103/RevModPhys.83.1523} {\bibfield  {journal} {\bibinfo
			{journal} {Rev. Mod. Phys.}\ }\textbf {\bibinfo {volume} {83}},\ \bibinfo
		{pages} {1523} (\bibinfo {year} {2011})}\BibitemShut {NoStop}%
	\bibitem [{\citenamefont {Goldman}\ and\ \citenamefont
		{Dalibard}(2014)}]{Goldman14}%
	\BibitemOpen
	\bibfield  {author} {\bibinfo {author} {\bibfnamefont {N.}~\bibnamefont
			{Goldman}}\ and\ \bibinfo {author} {\bibfnamefont {J.}~\bibnamefont
			{Dalibard}},\ }\href {https://doi.org/10.1103/PhysRevX.4.031027} {\bibfield
		{journal} {\bibinfo  {journal} {Phys. Rev. X}\ }\textbf {\bibinfo {volume}
			{4}},\ \bibinfo {pages} {031027} (\bibinfo {year} {2014})}\BibitemShut
	{NoStop}%
	\bibitem [{\citenamefont {Goldman}\ \emph {et~al.}(2016)\citenamefont
		{Goldman}, \citenamefont {Budich},\ and\ \citenamefont {Zoller}}]{Goldman16}%
	\BibitemOpen
	\bibfield  {author} {\bibinfo {author} {\bibfnamefont {N.}~\bibnamefont
			{Goldman}}, \bibinfo {author} {\bibfnamefont {J.~C.}\ \bibnamefont
			{Budich}},\ and\ \bibinfo {author} {\bibfnamefont {P.}~\bibnamefont
			{Zoller}},\ }\href {https://doi.org/10.1038/nphys3803} {\bibfield  {journal}
		{\bibinfo  {journal} {Nature Physics}\ }\textbf {\bibinfo {volume} {12}},\
		\bibinfo {pages} {639} (\bibinfo {year} {2016})}\BibitemShut {NoStop}%
	\bibitem [{\citenamefont {Jotzu}\ \emph {et~al.}(2014)\citenamefont {Jotzu},
		\citenamefont {Messer}, \citenamefont {Desbuquois}, \citenamefont {Lebrat},
		\citenamefont {Uehlinger}, \citenamefont {Greif},\ and\ \citenamefont
		{Esslinger}}]{Jotzu14}%
	\BibitemOpen
	\bibfield  {author} {\bibinfo {author} {\bibfnamefont {G.}~\bibnamefont
			{Jotzu}}, \bibinfo {author} {\bibfnamefont {M.}~\bibnamefont {Messer}},
		\bibinfo {author} {\bibfnamefont {R.}~\bibnamefont {Desbuquois}}, \bibinfo
		{author} {\bibfnamefont {M.}~\bibnamefont {Lebrat}}, \bibinfo {author}
		{\bibfnamefont {T.}~\bibnamefont {Uehlinger}}, \bibinfo {author}
		{\bibfnamefont {D.}~\bibnamefont {Greif}},\ and\ \bibinfo {author}
		{\bibfnamefont {T.}~\bibnamefont {Esslinger}},\ }\href
	{https://doi.org/10.1038/nature13915} {\bibfield  {journal} {\bibinfo
			{journal} {Nature}\ }\textbf {\bibinfo {volume} {515}},\ \bibinfo {pages}
		{237} (\bibinfo {year} {2014})}\BibitemShut {NoStop}%
	\bibitem [{\citenamefont {Fl{\"a}schner}\ \emph {et~al.}(2016)\citenamefont
		{Fl{\"a}schner}, \citenamefont {Rem}, \citenamefont {Tarnowski},
		\citenamefont {Vogel}, \citenamefont {L{\"u}hmann}, \citenamefont
		{Sengstock},\ and\ \citenamefont {Weitenberg}}]{Flaschner16}%
	\BibitemOpen
	\bibfield  {author} {\bibinfo {author} {\bibfnamefont {N.}~\bibnamefont
			{Fl{\"a}schner}}, \bibinfo {author} {\bibfnamefont {B.~S.}\ \bibnamefont
			{Rem}}, \bibinfo {author} {\bibfnamefont {M.}~\bibnamefont {Tarnowski}},
		\bibinfo {author} {\bibfnamefont {D.}~\bibnamefont {Vogel}}, \bibinfo
		{author} {\bibfnamefont {D.-S.}\ \bibnamefont {L{\"u}hmann}}, \bibinfo
		{author} {\bibfnamefont {K.}~\bibnamefont {Sengstock}},\ and\ \bibinfo
		{author} {\bibfnamefont {C.}~\bibnamefont {Weitenberg}},\ }\href
	{https://doi.org/10.1126/science.aad4568} {\bibfield  {journal} {\bibinfo
			{journal} {Science}\ }\textbf {\bibinfo {volume} {352}},\ \bibinfo {pages}
		{1091} (\bibinfo {year} {2016})}\BibitemShut {NoStop}%
	\bibitem [{\citenamefont {Wintersperger}\ \emph {et~al.}(2020)\citenamefont
		{Wintersperger}, \citenamefont {Braun}, \citenamefont {Ünal}, \citenamefont
		{Eckardt}, \citenamefont {Liberto}, \citenamefont {Goldman}, \citenamefont
		{Bloch},\ and\ \citenamefont {Aidelsburger}}]{Wintersperger20}%
	\BibitemOpen
	\bibfield  {author} {\bibinfo {author} {\bibfnamefont {K.}~\bibnamefont
			{Wintersperger}}, \bibinfo {author} {\bibfnamefont {C.}~\bibnamefont
			{Braun}}, \bibinfo {author} {\bibfnamefont {F.~N.}\ \bibnamefont {Ünal}},
		\bibinfo {author} {\bibfnamefont {A.}~\bibnamefont {Eckardt}}, \bibinfo
		{author} {\bibfnamefont {M.~D.}\ \bibnamefont {Liberto}}, \bibinfo {author}
		{\bibfnamefont {N.}~\bibnamefont {Goldman}}, \bibinfo {author} {\bibfnamefont
			{I.}~\bibnamefont {Bloch}},\ and\ \bibinfo {author} {\bibfnamefont
			{M.}~\bibnamefont {Aidelsburger}},\ }\href
	{https://doi.org/10.1038/s41567-020-0949-y} {\bibfield  {journal} {\bibinfo
			{journal} {Nature Physics}\ }\textbf {\bibinfo {volume} {16}},\ \bibinfo
		{pages} {1058} (\bibinfo {year} {2020})}\BibitemShut {NoStop}%
	\bibitem [{\citenamefont {Cooper}\ \emph {et~al.}(2019)\citenamefont {Cooper},
		\citenamefont {Dalibard},\ and\ \citenamefont {Spielman}}]{Cooper19}%
	\BibitemOpen
	\bibfield  {author} {\bibinfo {author} {\bibfnamefont {N.~R.}\ \bibnamefont
			{Cooper}}, \bibinfo {author} {\bibfnamefont {J.}~\bibnamefont {Dalibard}},\
		and\ \bibinfo {author} {\bibfnamefont {I.~B.}\ \bibnamefont {Spielman}},\
	}\href {https://doi.org/10.1103/RevModPhys.91.015005} {\bibfield  {journal}
		{\bibinfo  {journal} {Rev. Mod. Phys.}\ }\textbf {\bibinfo {volume} {91}},\
		\bibinfo {pages} {015005} (\bibinfo {year} {2019})}\BibitemShut {NoStop}%
	\bibitem [{\citenamefont {Oka}\ and\ \citenamefont
		{Kitamura}(2019)}]{Takashi19}%
	\BibitemOpen
	\bibfield  {author} {\bibinfo {author} {\bibfnamefont {T.}~\bibnamefont
			{Oka}}\ and\ \bibinfo {author} {\bibfnamefont {S.}~\bibnamefont {Kitamura}},\
	}\href {https://doi.org/10.1146/annurev-conmatphys-031218-013423} {\bibfield
		{journal} {\bibinfo  {journal} {Annual Review of Condensed Matter Physics}\
		}\textbf {\bibinfo {volume} {10}},\ \bibinfo {pages} {387} (\bibinfo {year}
		{2019})}\BibitemShut {NoStop}%
	\bibitem [{\citenamefont {Harper}\ \emph {et~al.}(2020)\citenamefont {Harper},
		\citenamefont {Roy}, \citenamefont {Rudner},\ and\ \citenamefont
		{Sondhi}}]{Harper20}%
	\BibitemOpen
	\bibfield  {author} {\bibinfo {author} {\bibfnamefont {F.}~\bibnamefont
			{Harper}}, \bibinfo {author} {\bibfnamefont {R.}~\bibnamefont {Roy}},
		\bibinfo {author} {\bibfnamefont {M.~S.}\ \bibnamefont {Rudner}},\ and\
		\bibinfo {author} {\bibfnamefont {S.}~\bibnamefont {Sondhi}},\ }\href
	{https://doi.org/10.1146/annurev-conmatphys-031218-013721} {\bibfield
		{journal} {\bibinfo  {journal} {Annual Review of Condensed Matter Physics}\
		}\textbf {\bibinfo {volume} {11}},\ \bibinfo {pages} {345} (\bibinfo {year}
		{2020})}\BibitemShut {NoStop}%
	\bibitem [{\citenamefont {Kitagawa}\ \emph {et~al.}(2010)\citenamefont
		{Kitagawa}, \citenamefont {Berg}, \citenamefont {Rudner},\ and\ \citenamefont
		{Demler}}]{Kitagawa10}%
	\BibitemOpen
	\bibfield  {author} {\bibinfo {author} {\bibfnamefont {T.}~\bibnamefont
			{Kitagawa}}, \bibinfo {author} {\bibfnamefont {E.}~\bibnamefont {Berg}},
		\bibinfo {author} {\bibfnamefont {M.}~\bibnamefont {Rudner}},\ and\ \bibinfo
		{author} {\bibfnamefont {E.}~\bibnamefont {Demler}},\ }\href
	{https://doi.org/10.1103/PhysRevB.82.235114} {\bibfield  {journal} {\bibinfo
			{journal} {Phys. Rev. B}\ }\textbf {\bibinfo {volume} {82}},\ \bibinfo
		{pages} {235114} (\bibinfo {year} {2010})}\BibitemShut {NoStop}%
	\bibitem [{\citenamefont {Rudner}\ \emph {et~al.}(2013)\citenamefont {Rudner},
		\citenamefont {Lindner}, \citenamefont {Berg},\ and\ \citenamefont
		{Levin}}]{Levin13}%
	\BibitemOpen
	\bibfield  {author} {\bibinfo {author} {\bibfnamefont {M.~S.}\ \bibnamefont
			{Rudner}}, \bibinfo {author} {\bibfnamefont {N.~H.}\ \bibnamefont {Lindner}},
		\bibinfo {author} {\bibfnamefont {E.}~\bibnamefont {Berg}},\ and\ \bibinfo
		{author} {\bibfnamefont {M.}~\bibnamefont {Levin}},\ }\href
	{https://doi.org/10.1103/PhysRevX.3.031005} {\bibfield  {journal} {\bibinfo
			{journal} {Phys. Rev. X}\ }\textbf {\bibinfo {volume} {3}},\ \bibinfo {pages}
		{031005} (\bibinfo {year} {2013})}\BibitemShut {NoStop}%
	\bibitem [{\citenamefont {Nathan}\ and\ \citenamefont
		{Rudner}(2015)}]{Nathan15}%
	\BibitemOpen
	\bibfield  {author} {\bibinfo {author} {\bibfnamefont {F.}~\bibnamefont
			{Nathan}}\ and\ \bibinfo {author} {\bibfnamefont {M.~S.}\ \bibnamefont
			{Rudner}},\ }\href {https://doi.org/10.1088/1367-2630/17/12/125014}
	{\bibfield  {journal} {\bibinfo  {journal} {New Journal of Physics}\ }\textbf
		{\bibinfo {volume} {17}},\ \bibinfo {pages} {125014} (\bibinfo {year}
		{2015})}\BibitemShut {NoStop}%
	\bibitem [{\citenamefont {Gemelke}\ \emph {et~al.}(2005)\citenamefont
		{Gemelke}, \citenamefont {Sarajlic}, \citenamefont {Bidel}, \citenamefont
		{Hong},\ and\ \citenamefont {Chu}}]{Gemelke05}%
	\BibitemOpen
	\bibfield  {author} {\bibinfo {author} {\bibfnamefont {N.}~\bibnamefont
			{Gemelke}}, \bibinfo {author} {\bibfnamefont {E.}~\bibnamefont {Sarajlic}},
		\bibinfo {author} {\bibfnamefont {Y.}~\bibnamefont {Bidel}}, \bibinfo
		{author} {\bibfnamefont {S.}~\bibnamefont {Hong}},\ and\ \bibinfo {author}
		{\bibfnamefont {S.}~\bibnamefont {Chu}},\ }\href
	{https://doi.org/10.1103/PhysRevLett.95.170404} {\bibfield  {journal}
		{\bibinfo  {journal} {Phys. Rev. Lett.}\ }\textbf {\bibinfo {volume} {95}},\
		\bibinfo {pages} {170404} (\bibinfo {year} {2005})}\BibitemShut {NoStop}%
	\bibitem [{\citenamefont {Parker}\ \emph {et~al.}(2013)\citenamefont {Parker},
		\citenamefont {Ha},\ and\ \citenamefont {Chin}}]{Chin13}%
	\BibitemOpen
	\bibfield  {author} {\bibinfo {author} {\bibfnamefont {C.~V.}\ \bibnamefont
			{Parker}}, \bibinfo {author} {\bibfnamefont {L.-C.}\ \bibnamefont {Ha}},\
		and\ \bibinfo {author} {\bibfnamefont {C.}~\bibnamefont {Chin}},\ }\href
	{https://doi.org/10.1038/nphys2789} {\bibfield  {journal} {\bibinfo
			{journal} {Nature Physics}\ }\textbf {\bibinfo {volume} {9}},\ \bibinfo
		{pages} {769} (\bibinfo {year} {2013})}\BibitemShut {NoStop}%
	\bibitem [{\citenamefont {Ha}\ \emph {et~al.}(2015)\citenamefont {Ha},
		\citenamefont {Clark}, \citenamefont {Parker}, \citenamefont {Anderson},\
		and\ \citenamefont {Chin}}]{Chin15}%
	\BibitemOpen
	\bibfield  {author} {\bibinfo {author} {\bibfnamefont {L.-C.}\ \bibnamefont
			{Ha}}, \bibinfo {author} {\bibfnamefont {L.~W.}\ \bibnamefont {Clark}},
		\bibinfo {author} {\bibfnamefont {C.~V.}\ \bibnamefont {Parker}}, \bibinfo
		{author} {\bibfnamefont {B.~M.}\ \bibnamefont {Anderson}},\ and\ \bibinfo
		{author} {\bibfnamefont {C.}~\bibnamefont {Chin}},\ }\href
	{https://doi.org/10.1103/PhysRevLett.114.055301} {\bibfield  {journal}
		{\bibinfo  {journal} {Phys. Rev. Lett.}\ }\textbf {\bibinfo {volume} {114}},\
		\bibinfo {pages} {055301} (\bibinfo {year} {2015})}\BibitemShut {NoStop}%
	\bibitem [{\citenamefont {Clark}\ \emph {et~al.}(2016)\citenamefont {Clark},
		\citenamefont {Feng},\ and\ \citenamefont {Chin}}]{Chicago16}%
	\BibitemOpen
	\bibfield  {author} {\bibinfo {author} {\bibfnamefont {L.~W.}\ \bibnamefont
			{Clark}}, \bibinfo {author} {\bibfnamefont {L.}~\bibnamefont {Feng}},\ and\
		\bibinfo {author} {\bibfnamefont {C.}~\bibnamefont {Chin}},\ }\href
	{https://doi.org/10.1126/science.aaf9657} {\bibfield  {journal} {\bibinfo
			{journal} {Science}\ }\textbf {\bibinfo {volume} {354}},\ \bibinfo {pages}
		{606} (\bibinfo {year} {2016})}\BibitemShut {NoStop}%
	\bibitem [{\citenamefont {Feng}\ \emph {et~al.}(2018)\citenamefont {Feng},
		\citenamefont {Clark}, \citenamefont {Gaj},\ and\ \citenamefont
		{Chin}}]{Chin18}%
	\BibitemOpen
	\bibfield  {author} {\bibinfo {author} {\bibfnamefont {L.}~\bibnamefont
			{Feng}}, \bibinfo {author} {\bibfnamefont {L.~W.}\ \bibnamefont {Clark}},
		\bibinfo {author} {\bibfnamefont {A.}~\bibnamefont {Gaj}},\ and\ \bibinfo
		{author} {\bibfnamefont {C.}~\bibnamefont {Chin}},\ }\href
	{https://doi.org/10.1038/s41567-017-0011-x} {\bibfield  {journal} {\bibinfo
			{journal} {Nature Physics}\ }\textbf {\bibinfo {volume} {14}},\ \bibinfo
		{pages} {269} (\bibinfo {year} {2018})}\BibitemShut {NoStop}%
	\bibitem [{\citenamefont {Clark}\ \emph {et~al.}(2018)\citenamefont {Clark},
		\citenamefont {Anderson}, \citenamefont {Feng}, \citenamefont {Gaj},
		\citenamefont {Levin},\ and\ \citenamefont {Chin}}]{Chicago18}%
	\BibitemOpen
	\bibfield  {author} {\bibinfo {author} {\bibfnamefont {L.~W.}\ \bibnamefont
			{Clark}}, \bibinfo {author} {\bibfnamefont {B.~M.}\ \bibnamefont {Anderson}},
		\bibinfo {author} {\bibfnamefont {L.}~\bibnamefont {Feng}}, \bibinfo {author}
		{\bibfnamefont {A.}~\bibnamefont {Gaj}}, \bibinfo {author} {\bibfnamefont
			{K.}~\bibnamefont {Levin}},\ and\ \bibinfo {author} {\bibfnamefont
			{C.}~\bibnamefont {Chin}},\ }\href
	{https://doi.org/10.1103/PhysRevLett.121.030402} {\bibfield  {journal}
		{\bibinfo  {journal} {Phys. Rev. Lett.}\ }\textbf {\bibinfo {volume} {121}},\
		\bibinfo {pages} {030402} (\bibinfo {year} {2018})}\BibitemShut {NoStop}%
	\bibitem [{\citenamefont {Zheng}\ \emph {et~al.}(2014)\citenamefont {Zheng},
		\citenamefont {Liu}, \citenamefont {Miao}, \citenamefont {Chin},\ and\
		\citenamefont {Zhai}}]{Zhai14}%
	\BibitemOpen
	\bibfield  {author} {\bibinfo {author} {\bibfnamefont {W.}~\bibnamefont
			{Zheng}}, \bibinfo {author} {\bibfnamefont {B.}~\bibnamefont {Liu}}, \bibinfo
		{author} {\bibfnamefont {J.}~\bibnamefont {Miao}}, \bibinfo {author}
		{\bibfnamefont {C.}~\bibnamefont {Chin}},\ and\ \bibinfo {author}
		{\bibfnamefont {H.}~\bibnamefont {Zhai}},\ }\href
	{https://doi.org/10.1103/PhysRevLett.113.155303} {\bibfield  {journal}
		{\bibinfo  {journal} {Phys. Rev. Lett.}\ }\textbf {\bibinfo {volume} {113}},\
		\bibinfo {pages} {155303} (\bibinfo {year} {2014})}\BibitemShut {NoStop}%
	\bibitem [{\citenamefont {Wirth}\ \emph {et~al.}(2010)\citenamefont {Wirth},
		\citenamefont {{\"O}lschl{\"a}ger},\ and\ \citenamefont
		{Hemmerich}}]{Wirth10}%
	\BibitemOpen
	\bibfield  {author} {\bibinfo {author} {\bibfnamefont {G.}~\bibnamefont
			{Wirth}}, \bibinfo {author} {\bibfnamefont {M.}~\bibnamefont
			{{\"O}lschl{\"a}ger}},\ and\ \bibinfo {author} {\bibfnamefont
			{A.}~\bibnamefont {Hemmerich}},\ }\href {https://doi.org/10.1038/nphys1857}
	{\bibfield  {journal} {\bibinfo  {journal} {Nature Physics}\ }\textbf
		{\bibinfo {volume} {7}},\ \bibinfo {pages} {147} (\bibinfo {year}
		{2010})}\BibitemShut {NoStop}%
	\bibitem [{\citenamefont {Soltan-Panahi}\ \emph {et~al.}(2011)\citenamefont
		{Soltan-Panahi}, \citenamefont {Lühmann}, \citenamefont {Struck},
		\citenamefont {Windpassinger},\ and\ \citenamefont
		{Sengstock}}]{SoltanPanahi11}%
	\BibitemOpen
	\bibfield  {author} {\bibinfo {author} {\bibfnamefont {P.}~\bibnamefont
			{Soltan-Panahi}}, \bibinfo {author} {\bibfnamefont {D.-S.}\ \bibnamefont
			{Lühmann}}, \bibinfo {author} {\bibfnamefont {J.}~\bibnamefont {Struck}},
		\bibinfo {author} {\bibfnamefont {P.}~\bibnamefont {Windpassinger}},\ and\
		\bibinfo {author} {\bibfnamefont {K.}~\bibnamefont {Sengstock}},\ }\href
	{https://doi.org/10.1038/nphys2128} {\bibfield  {journal} {\bibinfo
			{journal} {Nature Physics}\ }\textbf {\bibinfo {volume} {8}},\ \bibinfo
		{pages} {71} (\bibinfo {year} {2011})}\BibitemShut {NoStop}%
	\bibitem [{\citenamefont {Lewenstein}\ and\ \citenamefont
		{Liu}(2011)}]{Vincent11}%
	\BibitemOpen
	\bibfield  {author} {\bibinfo {author} {\bibfnamefont {M.}~\bibnamefont
			{Lewenstein}}\ and\ \bibinfo {author} {\bibfnamefont {W.~V.}\ \bibnamefont
			{Liu}},\ }\href {https://doi.org/10.1038/nphys1894} {\bibfield  {journal}
		{\bibinfo  {journal} {Nature Physics}\ }\textbf {\bibinfo {volume} {7}},\
		\bibinfo {pages} {101} (\bibinfo {year} {2011})}\BibitemShut {NoStop}%
	\bibitem [{\citenamefont {Kock}\ \emph {et~al.}(2015)\citenamefont {Kock},
		\citenamefont {\"Olschl\"ager}, \citenamefont {Ewerbeck}, \citenamefont
		{Huang}, \citenamefont {Mathey},\ and\ \citenamefont {Hemmerich}}]{Kock15}%
	\BibitemOpen
	\bibfield  {author} {\bibinfo {author} {\bibfnamefont {T.}~\bibnamefont
			{Kock}}, \bibinfo {author} {\bibfnamefont {M.}~\bibnamefont
			{\"Olschl\"ager}}, \bibinfo {author} {\bibfnamefont {A.}~\bibnamefont
			{Ewerbeck}}, \bibinfo {author} {\bibfnamefont {W.-M.}\ \bibnamefont {Huang}},
		\bibinfo {author} {\bibfnamefont {L.}~\bibnamefont {Mathey}},\ and\ \bibinfo
		{author} {\bibfnamefont {A.}~\bibnamefont {Hemmerich}},\ }\href
	{https://doi.org/10.1103/PhysRevLett.114.115301} {\bibfield  {journal}
		{\bibinfo  {journal} {Phys. Rev. Lett.}\ }\textbf {\bibinfo {volume} {114}},\
		\bibinfo {pages} {115301} (\bibinfo {year} {2015})}\BibitemShut {NoStop}%
	\bibitem [{\citenamefont {Kock}\ \emph {et~al.}(2016)\citenamefont {Kock},
		\citenamefont {Hippler}, \citenamefont {Ewerbeck},\ and\ \citenamefont
		{Hemmerich}}]{Kock16}%
	\BibitemOpen
	\bibfield  {author} {\bibinfo {author} {\bibfnamefont {T.}~\bibnamefont
			{Kock}}, \bibinfo {author} {\bibfnamefont {C.}~\bibnamefont {Hippler}},
		\bibinfo {author} {\bibfnamefont {A.}~\bibnamefont {Ewerbeck}},\ and\
		\bibinfo {author} {\bibfnamefont {A.}~\bibnamefont {Hemmerich}},\ }\href
	{https://doi.org/10.1088/0953-4075/49/4/042001} {\bibfield  {journal}
		{\bibinfo  {journal} {Journal of Physics B}\ }\textbf {\bibinfo {volume}
			{49}},\ \bibinfo {pages} {042001} (\bibinfo {year} {2016})}\BibitemShut
	{NoStop}%
	\bibitem [{\citenamefont {Li}\ and\ \citenamefont {Liu}(2016)}]{Li16}%
	\BibitemOpen
	\bibfield  {author} {\bibinfo {author} {\bibfnamefont {X.}~\bibnamefont
			{Li}}\ and\ \bibinfo {author} {\bibfnamefont {W.~V.}\ \bibnamefont {Liu}},\
	}\href {https://doi.org/10.1088/0034-4885/79/11/116401} {\bibfield  {journal}
		{\bibinfo  {journal} {Reports on Progress in Physics}\ }\textbf {\bibinfo
			{volume} {79}},\ \bibinfo {pages} {116401} (\bibinfo {year}
		{2016})}\BibitemShut {NoStop}%
	\bibitem [{\citenamefont {Shirley}(1965)}]{Shirley65}%
	\BibitemOpen
	\bibfield  {author} {\bibinfo {author} {\bibfnamefont {J.~H.}\ \bibnamefont
			{Shirley}},\ }\href {https://doi.org/10.1103/PhysRev.138.B979} {\bibfield
		{journal} {\bibinfo  {journal} {Phys. Rev.}\ }\textbf {\bibinfo {volume}
			{138}},\ \bibinfo {pages} {B979} (\bibinfo {year} {1965})}\BibitemShut
	{NoStop}%
	\bibitem [{\citenamefont {Sambe}(1973)}]{Sambe73}%
	\BibitemOpen
	\bibfield  {author} {\bibinfo {author} {\bibfnamefont {H.}~\bibnamefont
			{Sambe}},\ }\href {https://doi.org/10.1103/PhysRevA.7.2203} {\bibfield
		{journal} {\bibinfo  {journal} {Phys. Rev. A}\ }\textbf {\bibinfo {volume}
			{7}},\ \bibinfo {pages} {2203} (\bibinfo {year} {1973})}\BibitemShut
	{NoStop}%
	\bibitem [{\citenamefont {Dalfovo}\ \emph {et~al.}(1999)\citenamefont
		{Dalfovo}, \citenamefont {Giorgini}, \citenamefont {Pitaevskii},\ and\
		\citenamefont {Stringari}}]{Dalfovo99}%
	\BibitemOpen
	\bibfield  {author} {\bibinfo {author} {\bibfnamefont {F.}~\bibnamefont
			{Dalfovo}}, \bibinfo {author} {\bibfnamefont {S.}~\bibnamefont {Giorgini}},
		\bibinfo {author} {\bibfnamefont {L.~P.}\ \bibnamefont {Pitaevskii}},\ and\
		\bibinfo {author} {\bibfnamefont {S.}~\bibnamefont {Stringari}},\ }\href
	{https://doi.org/10.1103/RevModPhys.71.463} {\bibfield  {journal} {\bibinfo
			{journal} {Rev. Mod. Phys.}\ }\textbf {\bibinfo {volume} {71}},\ \bibinfo
		{pages} {463} (\bibinfo {year} {1999})}\BibitemShut {NoStop}%
	\bibitem [{\citenamefont {Pethick}\ and\ \citenamefont
		{Smith}(2008)}]{pethick08}%
	\BibitemOpen
	\bibfield  {author} {\bibinfo {author} {\bibfnamefont {C.~J.}\ \bibnamefont
			{Pethick}}\ and\ \bibinfo {author} {\bibfnamefont {H.}~\bibnamefont
			{Smith}},\ }\href {https://doi.org/10.1017/CBO9780511802850} {\emph {\bibinfo
			{title} {Bose–Einstein Condensation in Dilute Gases}}},\ \bibinfo {edition}
	{2nd}\ ed.\ (\bibinfo  {publisher} {Cambridge University Press},\ \bibinfo
	{year} {2008})\BibitemShut {NoStop}%
	\bibitem [{\citenamefont {Chen}(2020)}]{Chen20}%
	\BibitemOpen
	\bibfield  {author} {\bibinfo {author} {\bibfnamefont {H.}~\bibnamefont
			{Chen}},\ }\href {https://doi.org/10.1103/PhysRevA.101.063601} {\bibfield
		{journal} {\bibinfo  {journal} {Phys. Rev. A}\ }\textbf {\bibinfo {volume}
			{101}},\ \bibinfo {pages} {063601} (\bibinfo {year} {2020})}\BibitemShut
	{NoStop}%
	\bibitem [{\citenamefont {Nozi\'eres}(1995)}]{nozieres95}%
	\BibitemOpen
	\bibfield  {author} {\bibinfo {author} {\bibfnamefont {P.}~\bibnamefont
			{Nozi\'eres}},\ }\bibinfo {title} {Some comments on bose–einstein
		condensation},\ in\ \href {https://doi.org/10.1017/CBO9780511524240.004}
	{\emph {\bibinfo {booktitle} {Bose-Einstein Condensation}}},\ \bibinfo
	{editor} {edited by\ \bibinfo {editor} {\bibfnamefont {A.}~\bibnamefont
			{Griffin}}, \bibinfo {editor} {\bibfnamefont {D.~W.}\ \bibnamefont {Snoke}},\
		and\ \bibinfo {editor} {\bibfnamefont {S.}~\bibnamefont {Stringari}}}\
	(\bibinfo  {publisher} {Cambridge University Press},\ \bibinfo {year}
	{1995})\BibitemShut {NoStop}%
	\bibitem [{\citenamefont {Chen}\ and\ \citenamefont {Xie}(2018)}]{Chen18}%
	\BibitemOpen
	\bibfield  {author} {\bibinfo {author} {\bibfnamefont {H.}~\bibnamefont
			{Chen}}\ and\ \bibinfo {author} {\bibfnamefont {X.~C.}\ \bibnamefont {Xie}},\
	}\href {https://doi.org/10.1103/PhysRevA.98.053611} {\bibfield  {journal}
		{\bibinfo  {journal} {Phys. Rev. A}\ }\textbf {\bibinfo {volume} {98}},\
		\bibinfo {pages} {053611} (\bibinfo {year} {2018})}\BibitemShut {NoStop}%
	\bibitem [{\citenamefont {Xiao}\ \emph {et~al.}(2005)\citenamefont {Xiao},
		\citenamefont {Shi},\ and\ \citenamefont {Niu}}]{Xiao05}%
	\BibitemOpen
	\bibfield  {author} {\bibinfo {author} {\bibfnamefont {D.}~\bibnamefont
			{Xiao}}, \bibinfo {author} {\bibfnamefont {J.}~\bibnamefont {Shi}},\ and\
		\bibinfo {author} {\bibfnamefont {Q.}~\bibnamefont {Niu}},\ }\href
	{https://doi.org/10.1103/PhysRevLett.95.137204} {\bibfield  {journal}
		{\bibinfo  {journal} {Phys. Rev. Lett.}\ }\textbf {\bibinfo {volume} {95}},\
		\bibinfo {pages} {137204} (\bibinfo {year} {2005})}\BibitemShut {NoStop}%
	\bibitem [{\citenamefont {Thonhauser}\ \emph {et~al.}(2005)\citenamefont
		{Thonhauser}, \citenamefont {Ceresoli}, \citenamefont {Vanderbilt},\ and\
		\citenamefont {Resta}}]{Thonhauser05}%
	\BibitemOpen
	\bibfield  {author} {\bibinfo {author} {\bibfnamefont {T.}~\bibnamefont
			{Thonhauser}}, \bibinfo {author} {\bibfnamefont {D.}~\bibnamefont
			{Ceresoli}}, \bibinfo {author} {\bibfnamefont {D.}~\bibnamefont
			{Vanderbilt}},\ and\ \bibinfo {author} {\bibfnamefont {R.}~\bibnamefont
			{Resta}},\ }\href {https://doi.org/10.1103/PhysRevLett.95.137205} {\bibfield
		{journal} {\bibinfo  {journal} {Phys. Rev. Lett.}\ }\textbf {\bibinfo
			{volume} {95}},\ \bibinfo {pages} {137205} (\bibinfo {year}
		{2005})}\BibitemShut {NoStop}%
	\bibitem [{\citenamefont {Ceresoli}\ \emph {et~al.}(2006)\citenamefont
		{Ceresoli}, \citenamefont {Thonhauser}, \citenamefont {Vanderbilt},\ and\
		\citenamefont {Resta}}]{Ceresoli06}%
	\BibitemOpen
	\bibfield  {author} {\bibinfo {author} {\bibfnamefont {D.}~\bibnamefont
			{Ceresoli}}, \bibinfo {author} {\bibfnamefont {T.}~\bibnamefont
			{Thonhauser}}, \bibinfo {author} {\bibfnamefont {D.}~\bibnamefont
			{Vanderbilt}},\ and\ \bibinfo {author} {\bibfnamefont {R.}~\bibnamefont
			{Resta}},\ }\href {https://doi.org/10.1103/PhysRevB.74.024408} {\bibfield
		{journal} {\bibinfo  {journal} {Phys. Rev. B}\ }\textbf {\bibinfo {volume}
			{74}},\ \bibinfo {pages} {024408} (\bibinfo {year} {2006})}\BibitemShut
	{NoStop}%
	\bibitem [{\citenamefont {Shi}\ \emph {et~al.}(2007)\citenamefont {Shi},
		\citenamefont {Vignale}, \citenamefont {Xiao},\ and\ \citenamefont
		{Niu}}]{Shi07}%
	\BibitemOpen
	\bibfield  {author} {\bibinfo {author} {\bibfnamefont {J.}~\bibnamefont
			{Shi}}, \bibinfo {author} {\bibfnamefont {G.}~\bibnamefont {Vignale}},
		\bibinfo {author} {\bibfnamefont {D.}~\bibnamefont {Xiao}},\ and\ \bibinfo
		{author} {\bibfnamefont {Q.}~\bibnamefont {Niu}},\ }\href
	{https://doi.org/10.1103/PhysRevLett.99.197202} {\bibfield  {journal}
		{\bibinfo  {journal} {Phys. Rev. Lett.}\ }\textbf {\bibinfo {volume} {99}},\
		\bibinfo {pages} {197202} (\bibinfo {year} {2007})}\BibitemShut {NoStop}%
	\bibitem [{\citenamefont {Xiao}\ \emph {et~al.}(2010)\citenamefont {Xiao},
		\citenamefont {Chang},\ and\ \citenamefont {Niu}}]{Xiao10}%
	\BibitemOpen
	\bibfield  {author} {\bibinfo {author} {\bibfnamefont {D.}~\bibnamefont
			{Xiao}}, \bibinfo {author} {\bibfnamefont {M.-C.}\ \bibnamefont {Chang}},\
		and\ \bibinfo {author} {\bibfnamefont {Q.}~\bibnamefont {Niu}},\ }\href
	{https://doi.org/10.1103/RevModPhys.82.1959} {\bibfield  {journal} {\bibinfo
			{journal} {Rev. Mod. Phys.}\ }\textbf {\bibinfo {volume} {82}},\ \bibinfo
		{pages} {1959} (\bibinfo {year} {2010})}\BibitemShut {NoStop}%
	\bibitem [{\citenamefont {Thonhauser}(2011)}]{Thonhauser11}%
	\BibitemOpen
	\bibfield  {author} {\bibinfo {author} {\bibfnamefont {T.}~\bibnamefont
			{Thonhauser}},\ }\href {https://doi.org/10.1142/S0217979211058912} {\bibfield
		{journal} {\bibinfo  {journal} {Int. J. Mod. Phys. B}\ }\textbf {\bibinfo
			{volume} {25}},\ \bibinfo {pages} {1429} (\bibinfo {year}
		{2011})}\BibitemShut {NoStop}%
	\bibitem [{\citenamefont {Bohm}\ \emph {et~al.}(2003)\citenamefont {Bohm},
		\citenamefont {Mostafazadeh}, \citenamefont {Koizumi}, \citenamefont {Niu},\
		and\ \citenamefont {Zwanziger}}]{Bohm03}%
	\BibitemOpen
	\bibfield  {author} {\bibinfo {author} {\bibfnamefont {A.}~\bibnamefont
			{Bohm}}, \bibinfo {author} {\bibfnamefont {A.}~\bibnamefont {Mostafazadeh}},
		\bibinfo {author} {\bibfnamefont {H.}~\bibnamefont {Koizumi}}, \bibinfo
		{author} {\bibfnamefont {Q.}~\bibnamefont {Niu}},\ and\ \bibinfo {author}
		{\bibfnamefont {J.}~\bibnamefont {Zwanziger}},\ }\href
	{https://doi.org/10.1007/978-3-662-10333-3} {\emph {\bibinfo {title} {The
				Geometric Phase in Quantum Systems}}}\ (\bibinfo  {publisher} {Springer
		Berlin Heidelberg},\ \bibinfo {year} {2003})\BibitemShut {NoStop}%
	\bibitem [{\citenamefont {Wilczek}\ and\ \citenamefont
		{Shapere}(1989)}]{Wilczek89}%
	\BibitemOpen
	\bibfield  {author} {\bibinfo {author} {\bibfnamefont {F.}~\bibnamefont
			{Wilczek}}\ and\ \bibinfo {author} {\bibfnamefont {A.}~\bibnamefont
			{Shapere}},\ }\href {https://doi.org/10.1142/0613} {\emph {\bibinfo {title}
			{Geometric Phases in Physics}}}\ (\bibinfo  {publisher} {{WORLD}
		{SCIENTIFIC}},\ \bibinfo {year} {1989})\BibitemShut {NoStop}%
	\bibitem [{\citenamefont {Fukui}\ \emph {et~al.}(2005)\citenamefont {Fukui},
		\citenamefont {Hatsugai},\ and\ \citenamefont {Suzuki}}]{Takahiro05}%
	\BibitemOpen
	\bibfield  {author} {\bibinfo {author} {\bibfnamefont {T.}~\bibnamefont
			{Fukui}}, \bibinfo {author} {\bibfnamefont {Y.}~\bibnamefont {Hatsugai}},\
		and\ \bibinfo {author} {\bibfnamefont {H.}~\bibnamefont {Suzuki}},\ }\href
	{https://doi.org/10.1143/JPSJ.74.1674} {\bibfield  {journal} {\bibinfo
			{journal} {Journal of the Physical Society of Japan}\ }\textbf {\bibinfo
			{volume} {74}},\ \bibinfo {pages} {1674} (\bibinfo {year}
		{2005})}\BibitemShut {NoStop}%
	\bibitem [{\citenamefont {Dauphin}\ and\ \citenamefont
		{Goldman}(2013)}]{Goldman13}%
	\BibitemOpen
	\bibfield  {author} {\bibinfo {author} {\bibfnamefont {A.}~\bibnamefont
			{Dauphin}}\ and\ \bibinfo {author} {\bibfnamefont {N.}~\bibnamefont
			{Goldman}},\ }\href {https://doi.org/10.1103/PhysRevLett.111.135302}
	{\bibfield  {journal} {\bibinfo  {journal} {Phys. Rev. Lett.}\ }\textbf
		{\bibinfo {volume} {111}},\ \bibinfo {pages} {135302} (\bibinfo {year}
		{2013})}\BibitemShut {NoStop}%
	\bibitem [{\citenamefont {Liu}\ \emph {et~al.}(2013)\citenamefont {Liu},
		\citenamefont {Law}, \citenamefont {Ng},\ and\ \citenamefont {Lee}}]{Liu13}%
	\BibitemOpen
	\bibfield  {author} {\bibinfo {author} {\bibfnamefont {X.-J.}\ \bibnamefont
			{Liu}}, \bibinfo {author} {\bibfnamefont {K.~T.}\ \bibnamefont {Law}},
		\bibinfo {author} {\bibfnamefont {T.~K.}\ \bibnamefont {Ng}},\ and\ \bibinfo
		{author} {\bibfnamefont {P.~A.}\ \bibnamefont {Lee}},\ }\href
	{https://doi.org/10.1103/PhysRevLett.111.120402} {\bibfield  {journal}
		{\bibinfo  {journal} {Phys. Rev. Lett.}\ }\textbf {\bibinfo {volume} {111}},\
		\bibinfo {pages} {120402} (\bibinfo {year} {2013})}\BibitemShut {NoStop}%
	\bibitem [{\citenamefont {Aidelsburger}\ \emph {et~al.}(2015)\citenamefont
		{Aidelsburger}, \citenamefont {Lohse}, \citenamefont {Schweizer},
		\citenamefont {Atala}, \citenamefont {Barreiro}, \citenamefont
		{Nascimb\`ene}, \citenamefont {Cooper}, \citenamefont {Bloch},\ and\
		\citenamefont {Goldman}}]{Aidelsburger15}%
	\BibitemOpen
	\bibfield  {author} {\bibinfo {author} {\bibfnamefont {M.}~\bibnamefont
			{Aidelsburger}}, \bibinfo {author} {\bibfnamefont {M.}~\bibnamefont {Lohse}},
		\bibinfo {author} {\bibfnamefont {C.}~\bibnamefont {Schweizer}}, \bibinfo
		{author} {\bibfnamefont {M.}~\bibnamefont {Atala}}, \bibinfo {author}
		{\bibfnamefont {J.~T.}\ \bibnamefont {Barreiro}}, \bibinfo {author}
		{\bibfnamefont {S.}~\bibnamefont {Nascimb\`ene}}, \bibinfo {author}
		{\bibfnamefont {N.~R.}\ \bibnamefont {Cooper}}, \bibinfo {author}
		{\bibfnamefont {I.}~\bibnamefont {Bloch}},\ and\ \bibinfo {author}
		{\bibfnamefont {N.}~\bibnamefont {Goldman}},\ }\href
	{https://doi.org/10.1038/nphys3171} {\bibfield  {journal} {\bibinfo
			{journal} {Nature Physics}\ }\textbf {\bibinfo {volume} {11}},\ \bibinfo
		{pages} {162} (\bibinfo {year} {2015})}\BibitemShut {NoStop}%
	\bibitem [{\citenamefont {Resta}\ and\ \citenamefont
		{Vanderbilt}(2007)}]{Resta2007}%
	\BibitemOpen
	\bibfield  {author} {\bibinfo {author} {\bibfnamefont {R.}~\bibnamefont
			{Resta}}\ and\ \bibinfo {author} {\bibfnamefont {D.}~\bibnamefont
			{Vanderbilt}},\ }\bibinfo {title} {Theory of polarization: A modern
		approach},\ in\ \href {https://doi.org/10.1007/978-3-540-34591-6_2} {\emph
		{\bibinfo {booktitle} {Physics of Ferroelectrics: A Modern Perspective}}}\
	(\bibinfo  {publisher} {Springer Berlin Heidelberg},\ \bibinfo {address}
	{Berlin, Heidelberg},\ \bibinfo {year} {2007})\BibitemShut {NoStop}%
	\bibitem [{\citenamefont {Vanderbilt}(2018)}]{Vanderbilt18}%
	\BibitemOpen
	\bibfield  {author} {\bibinfo {author} {\bibfnamefont {D.}~\bibnamefont
			{Vanderbilt}},\ }\href {https://doi.org/10.1017/9781316662205} {\emph
		{\bibinfo {title} {Berry Phases in Electronic Structure Theory}}}\ (\bibinfo
	{publisher} {Cambridge University Press},\ \bibinfo {year}
	{2018})\BibitemShut {NoStop}%
	\bibitem [{\citenamefont {Huang}\ and\ \citenamefont {Liu}(2020)}]{Liu20}%
	\BibitemOpen
	\bibfield  {author} {\bibinfo {author} {\bibfnamefont {B.}~\bibnamefont
			{Huang}}\ and\ \bibinfo {author} {\bibfnamefont {W.~V.}\ \bibnamefont
			{Liu}},\ }\href {https://doi.org/10.1103/PhysRevLett.124.216601} {\bibfield
		{journal} {\bibinfo  {journal} {Phys. Rev. Lett.}\ }\textbf {\bibinfo
			{volume} {124}},\ \bibinfo {pages} {216601} (\bibinfo {year}
		{2020})}\BibitemShut {NoStop}%
	\bibitem [{\citenamefont {Nakajima}\ \emph {et~al.}(2016)\citenamefont
		{Nakajima}, \citenamefont {Tomita}, \citenamefont {Taie}, \citenamefont
		{Ichinose}, \citenamefont {Ozawa}, \citenamefont {Wang}, \citenamefont
		{Troyer},\ and\ \citenamefont {Takahashi}}]{Nakajima16}%
	\BibitemOpen
	\bibfield  {author} {\bibinfo {author} {\bibfnamefont {S.}~\bibnamefont
			{Nakajima}}, \bibinfo {author} {\bibfnamefont {T.}~\bibnamefont {Tomita}},
		\bibinfo {author} {\bibfnamefont {S.}~\bibnamefont {Taie}}, \bibinfo {author}
		{\bibfnamefont {T.}~\bibnamefont {Ichinose}}, \bibinfo {author}
		{\bibfnamefont {H.}~\bibnamefont {Ozawa}}, \bibinfo {author} {\bibfnamefont
			{L.}~\bibnamefont {Wang}}, \bibinfo {author} {\bibfnamefont {M.}~\bibnamefont
			{Troyer}},\ and\ \bibinfo {author} {\bibfnamefont {Y.}~\bibnamefont
			{Takahashi}},\ }\href {https://doi.org/10.1038/nphys3622} {\bibfield
		{journal} {\bibinfo  {journal} {Nature Physics}\ }\textbf {\bibinfo {volume}
			{12}},\ \bibinfo {pages} {296} (\bibinfo {year} {2016})}\BibitemShut
	{NoStop}%
	\bibitem [{\citenamefont {Lohse}\ \emph {et~al.}(2016)\citenamefont {Lohse},
		\citenamefont {Schweizer}, \citenamefont {Zilberberg}, \citenamefont
		{Aidelsburger},\ and\ \citenamefont {Bloch}}]{Lohse16}%
	\BibitemOpen
	\bibfield  {author} {\bibinfo {author} {\bibfnamefont {M.}~\bibnamefont
			{Lohse}}, \bibinfo {author} {\bibfnamefont {C.}~\bibnamefont {Schweizer}},
		\bibinfo {author} {\bibfnamefont {O.}~\bibnamefont {Zilberberg}}, \bibinfo
		{author} {\bibfnamefont {M.}~\bibnamefont {Aidelsburger}},\ and\ \bibinfo
		{author} {\bibfnamefont {I.}~\bibnamefont {Bloch}},\ }\href
	{https://doi.org/10.1038/nphys3584} {\bibfield  {journal} {\bibinfo
			{journal} {Nature Physics}\ }\textbf {\bibinfo {volume} {12}},\ \bibinfo
		{pages} {350} (\bibinfo {year} {2016})}\BibitemShut {NoStop}%
	\bibitem [{\citenamefont {Marzari}\ and\ \citenamefont
		{Vanderbilt}(1997)}]{Vanderbilt97}%
	\BibitemOpen
	\bibfield  {author} {\bibinfo {author} {\bibfnamefont {N.}~\bibnamefont
			{Marzari}}\ and\ \bibinfo {author} {\bibfnamefont {D.}~\bibnamefont
			{Vanderbilt}},\ }\href {https://doi.org/10.1103/PhysRevB.56.12847} {\bibfield
		{journal} {\bibinfo  {journal} {Phys. Rev. B}\ }\textbf {\bibinfo {volume}
			{56}},\ \bibinfo {pages} {12847} (\bibinfo {year} {1997})}\BibitemShut
	{NoStop}%
	\bibitem [{\citenamefont {Souza}\ \emph {et~al.}(2001)\citenamefont {Souza},
		\citenamefont {Marzari},\ and\ \citenamefont {Vanderbilt}}]{Vanderbilt01}%
	\BibitemOpen
	\bibfield  {author} {\bibinfo {author} {\bibfnamefont {I.}~\bibnamefont
			{Souza}}, \bibinfo {author} {\bibfnamefont {N.}~\bibnamefont {Marzari}},\
		and\ \bibinfo {author} {\bibfnamefont {D.}~\bibnamefont {Vanderbilt}},\
	}\href {https://doi.org/10.1103/PhysRevB.65.035109} {\bibfield  {journal}
		{\bibinfo  {journal} {Phys. Rev. B}\ }\textbf {\bibinfo {volume} {65}},\
		\bibinfo {pages} {035109} (\bibinfo {year} {2001})}\BibitemShut {NoStop}%
	\bibitem [{\citenamefont {Marzari}\ \emph {et~al.}(2012)\citenamefont
		{Marzari}, \citenamefont {Mostofi}, \citenamefont {Yates}, \citenamefont
		{Souza},\ and\ \citenamefont {Vanderbilt}}]{Vanderbilt12}%
	\BibitemOpen
	\bibfield  {author} {\bibinfo {author} {\bibfnamefont {N.}~\bibnamefont
			{Marzari}}, \bibinfo {author} {\bibfnamefont {A.~A.}\ \bibnamefont
			{Mostofi}}, \bibinfo {author} {\bibfnamefont {J.~R.}\ \bibnamefont {Yates}},
		\bibinfo {author} {\bibfnamefont {I.}~\bibnamefont {Souza}},\ and\ \bibinfo
		{author} {\bibfnamefont {D.}~\bibnamefont {Vanderbilt}},\ }\href
	{https://doi.org/10.1103/RevModPhys.84.1419} {\bibfield  {journal} {\bibinfo
			{journal} {Rev. Mod. Phys.}\ }\textbf {\bibinfo {volume} {84}},\ \bibinfo
		{pages} {1419} (\bibinfo {year} {2012})}\BibitemShut {NoStop}%
	\bibitem [{\citenamefont {Mostofi}\ \emph {et~al.}(2008)\citenamefont
		{Mostofi}, \citenamefont {Yates}, \citenamefont {Lee}, \citenamefont {Souza},
		\citenamefont {Vanderbilt},\ and\ \citenamefont {Marzari}}]{Mostofi08}%
	\BibitemOpen
	\bibfield  {author} {\bibinfo {author} {\bibfnamefont {A.~A.}\ \bibnamefont
			{Mostofi}}, \bibinfo {author} {\bibfnamefont {J.~R.}\ \bibnamefont {Yates}},
		\bibinfo {author} {\bibfnamefont {Y.-S.}\ \bibnamefont {Lee}}, \bibinfo
		{author} {\bibfnamefont {I.}~\bibnamefont {Souza}}, \bibinfo {author}
		{\bibfnamefont {D.}~\bibnamefont {Vanderbilt}},\ and\ \bibinfo {author}
		{\bibfnamefont {N.}~\bibnamefont {Marzari}},\ }\href
	{https://doi.org/https://doi.org/10.1016/j.cpc.2007.11.016} {\bibfield
		{journal} {\bibinfo  {journal} {Computer Physics Communications}\ }\textbf
		{\bibinfo {volume} {178}},\ \bibinfo {pages} {685 } (\bibinfo {year}
		{2008})}\BibitemShut {NoStop}%
	\bibitem [{\citenamefont {Kivelson}(1982)}]{Kivelson82}%
	\BibitemOpen
	\bibfield  {author} {\bibinfo {author} {\bibfnamefont {S.}~\bibnamefont
			{Kivelson}},\ }\href {https://doi.org/10.1103/PhysRevB.26.4269} {\bibfield
		{journal} {\bibinfo  {journal} {Phys. Rev. B}\ }\textbf {\bibinfo {volume}
			{26}},\ \bibinfo {pages} {4269} (\bibinfo {year} {1982})}\BibitemShut
	{NoStop}%
	\bibitem [{\citenamefont {Resta}(1998)}]{Resta98}%
	\BibitemOpen
	\bibfield  {author} {\bibinfo {author} {\bibfnamefont {R.}~\bibnamefont
			{Resta}},\ }\href {https://doi.org/10.1103/PhysRevLett.80.1800} {\bibfield
		{journal} {\bibinfo  {journal} {Phys. Rev. Lett.}\ }\textbf {\bibinfo
			{volume} {80}},\ \bibinfo {pages} {1800} (\bibinfo {year}
		{1998})}\BibitemShut {NoStop}%
\end{thebibliography}

%

\end{document}